\documentclass[aps,pre,showpacs,twocolumn]{revtex4-1}

\usepackage{amsmath,amssymb,amsfonts,hyperref}
\usepackage{graphicx,epsf}
\usepackage{xspace}
\usepackage{textcomp}
\usepackage{color}

\newif\ifhyper
\hypertrue
\ifhyper
\hypersetup{
  citecolor = {green},
  urlcolor = {blue} 
} 

\newcommand{\p}{\partial}

\newcommand{\vx}{\vec{x}}

\newcommand{\vv}{\vec{v}}
\newcommand{\vnabla}{\vec{\nabla}}

\newcommand{\vf}{\vec{f}\,}

\newcommand{\hx}{{\hat x}}
\newcommand{\hy}{{\hat y}}

\newcommand{\vq}{{\vec{q}}}

\newcommand{\vk}{{\vec{k}}}

\newcommand{\FRG}{NPRG\xspace}

\begin{document}

\title{Spatio-temporal velocity-velocity correlation function in fully developed  turbulence}

\author{L\'eonie Canet$^{1}$, Vincent Rossetto$^{1}$, Nicol\'as Wschebor$^{2}$, and Guillaume Balarac$^{3}$}

\affiliation{$^1$ Universit\'e Grenoble Alpes and CNRS, LPMMC, UMR 5493, 38042 Grenoble, France \\$^2$ Instituto de F\'isica, Facultad de Ingenier\'ia, Universidad de la Rep\'ublica, J.H.y Reissig 565, 11000 Montevideo, Uruguay \\$^3$  Universit\'e Grenoble Alpes and CNRS, LEGI, UMR 5519, 38042 Grenoble, France}%\\$^*$ email: leonie.canet@grenoble.cnrs.fr}

\begin{abstract}

Turbulence is an ubiquitous phenomenon in natural and industrial flows.
  Since the celebrated work of Kolmogorov in 1941, 
 understanding the statistical properties of fully developed turbulence has remained a major quest.
 In particular, deriving the properties of turbulent flows from a mesoscopic description, that is from  Navier-Stokes
 equation, has eluded most theoretical attempts.
 Here, we provide a theoretical prediction for the functional  {\it space and time} dependence of velocity-velocity
 correlation function of homogeneous and isotropic turbulence
 from the field theory associated to  Navier-Stokes equation with stochastic forcing.
 This prediction, which goes beyond Kolmogorov theory,
  is the analytical fixed-point solution of Non-Perturbative Renormalization Group flow equations, 
 which are exact in the limit of large wave-numbers.
 This solution is compared to two-point two-times correlation functions computed in direct numerical simulations.
  We obtain a remarkable agreement both in the inertial and in the dissipative ranges.

\end{abstract}

\maketitle

\section{Introduction}

Kolmogorov made a fundamental step in the understanding of the statistical properties of 
homogeneous and isotropic three-dimensional turbulence in his seminal K41 theory \cite{Kolmogorov41a,Kolmogorov41c}. 
 This theory shows that  energy, which is injected at a typical large (integral) length scale $L$ by the external stirring, is conserved across an inertial range of scales, through a constant-flux transfer mechanism (the energy cascade),  until it is dissipated by molecular viscosity at a small (Kolmogorov) length scale $\eta$.
  Assuming universality and scale invariance in the inertial range, one then deduce from dimensional 
 considerations  scaling predictions such as the power-law decay of the
   kinetic energy spectrum with the well-known -5/3 exponent.
 These predictions quite reliably describe  most experimental and numerical observations,
 at least for the energy spectrum and low-order structure functions (moments of equal-time velocity differences)  \cite{Frisch95}.
  However, despite many theoretical efforts, the derivation of these scaling predictions from  fundamental principles, 
 that is from Navier-Stokes (NS) equation for the fluid dynamics,
 is still  unsatisfactory  \cite{Frisch95}. Moreover,  deviations from K41 scalings are observed
  in experiments and numerical simulations and are large for high-order structure functions.
   These deviations are related to what is named ``intermittency'', which refers to the full-fledged
    complexity of turbulence beyond K41 theory \cite{Frisch95}. 
    Calculating intermittency effects from NS equations is a longstanding
    unsolved issue.

 In this work, we derive  analytical solutions of  fixed-point Non-Perturbative (or Functional)
  Renormalisation Group (\FRG)  equations associated with  NS equation. These equations are the exact leading behavior
 at large wave-numbers. 
 We obtain the full space and time dependence of correlation and response
  functions in the turbulent steady-state at distances smaller than the integral scale $L$.
 Its spatial Fourier transform  
   is found to take a form $\propto\exp(-\alpha k^2 t^2)$, where $k$ is the wave-number, $t$ the time interval, and $\alpha$ 
   a (non-universal) constant which could be calculated from the (numerical) solution of the full flow equations.
  Let us emphasize that the dependence in $tk$ is a noticeable result. 
   Indeed, scaling theory would imply that the correlation function depends on the scaling
    variable $tk^z$, where $z$ is the dynamical exponent, which is  $z=2/3$ for Navier-Stokes turbulence in $d=3$.
     This result shows that an effective exponent $z=1$ arises, which is a signature of violations of scale invariance,
    that is intermittency. It follows in particular that 
      the presented solution correctly  accounts for the ``sweeping effect'', which is imposed by  the large-scale motion
  on the Eulerian velocities at small scales \cite{Tennekes75}. 
  Indeed, the  energy spectrum  calculated as a function of the frequency displays a $\omega^{-5/3}$ decay,
  as observed in experiments or numerical simulations for Eulerian velocity correlations \cite{Frisch95,Chevillard05}.
  This is a non-trivial result
   from a theoretical point of view \cite{Yakhot89,Chen89,Nelkin89,Gotoh93} which reflects deviations from K41 scalings.
   
  On the other hand, at {\it coinciding times}, the violations to scale invariance are found to be sub-leading
   at large $k$, and are hence not captured by the exact equations at leading order 
    studied in this work. In particular, determining 
   corrections to the  $k^{-5/3}$ decay of the energy spectrum in the inertial range would require the study of sub-leading equations, which goes beyond the present work.
  However,  the behavior of this spectrum can be determined for scales in the dissipative range, beyond the Kolmogorov regime. 
  Several mostly empirical expressions were proposed to describe this behavior 
   \cite{Monin73}.  They all suggest ``an approximately exponential decay'' \cite{Nelkin89}.
  Our analytical solution  shows that a crossover occurs at a scale given by Taylor scale $\lambda$, from the $k^{-5/3}$ power-law in the inertial range to  a stretched exponential decay following $\propto\exp(-\hat \mu (\lambda k)^{2/3})$ in the dissipative range, where  $\hat \mu$ is a (non-universal) constant (which 
  could be computed by numerically integrating the full flow equations).
    
  In order to test these predictions, we perform direct numerical simulations of fully developed isotropic turbulence from NS
 equation,  recording in particular the time-dependence of the correlation function. The numerical solution precisely exhibits the Gaussian
  dependence in $kt$. Moreover, the behavior of the numerical energy spectrum
   in the dissipative range is in agreement with the predicted stretched exponential.
  
In summary, we provide  an analytical expression for the spatiotemporal correlation and response function,  
    accounting in particular for  intermittency corrections at finite time differences, and which is 
    confirmed by the numerical data.  This expression is directly derived from NS equation 
    without approximation for large wave-numbers.   This constitutes
  a major step in the theoretical understanding and modeling of isotropic and homogeneous  turbulence,
  and  opens promising perspectives
 for the calculation of higher-order correlation and structure functions for three-dimensional,  and also
  two-dimensional turbulence.

The paper is organized as follows. The principles of the \FRG formalism and main results obtained in \cite{Canet16}
are reviewed  in Sec.~\ref{SEC:NPRG}. Our starting point is the flow equation for the two-point functions 
in an asymptotic form which is exact at large wave-numbers.  We focus on the fixed point equations, 
  derive and analyze  their general solution  in Sec.~\ref{SEC:inertial}.
  We  study in particular the behavior of the solution in 
  the dissipative range in Sec.~\ref{SEC:dissipative}.
      We  briefly describe in Sec.~\ref{SEC:DNS} the direct numerical   simulations, 
    before stressing concluding remarks and perspectives.

\section{NPRG formalism}
\label{SEC:NPRG}

  The results presented in this work are based on the mesoscopic description of fluid dynamics
  embodied  in Navier-Stokes equation
\begin{equation}
 \partial_t \vv+ \vv \cdot \vnabla \vv=-\frac 1\rho 
\vnabla p +\nu \nabla^2 \vv +\vf \,.
\label{ns}
\end{equation}
In this equation,  the velocity field $\vv$, the pressure field $p$, and the 
 external stirring force $\vec f$ depend on the 
space-time coordinates $(t,\vx)$,  $\nu$  is the kinematic viscosity and 
$\rho$ the density of the fluid. This continuous hydrodynamical description is
 typically valid at scales much smaller than the Kolmogorov scale defined by
\begin{equation}
 \eta = \left(\frac{\nu^3}{\epsilon}\right)^{1/4}\, 
\label{eq:def-eta}
\end{equation}
where $\epsilon$ is the mean rate of injection of energy per unit mass. 
The presence of the external forcing $\vf$ in (\ref{ns})
   is necessary to sustain a stationary turbulent state.
 We consider incompressible flows, satisfying 
 $\vnabla \cdot \vv = 0$. 

  At characteristic distances much smaller than the integral scale, the statistical behavior of the velocity field
   is observed to be independent of the actual details of the forcing. Therefore,  one can conveniently  perform
  a statistical average over stochastic forcings peaked at scales of order $L$.
 These forcings are chosen Gaussian distributed, with a correlator
\begin{equation}
 \langle f_\alpha(t,\vx)f_\beta(t',\vx\,')\rangle=2 \delta_{\alpha\beta}\delta(t-t')N_{L}(|\vx-\vx\,'|)\, ,
 \label{eq:correlator}
\end{equation}
where the profile $N_{L}$ is peaked at the scale $L$.
  NS equation with stochastic forcing can then be cast into a field theory 
  following the standard Martin-Siggia-Rose-Janssen-de Dominicis formalism \cite{Martin73,Janssen76,Dominicis76}.
  
   Since universality and power-law behaviors are expected in the inertial range, 
   the Renormalization Group (RG) appears as a natural theoretical approach to study the NS field theory,
   and to calculate its scaling properties \cite{zinnjustin89}.
  However, applying perturbative RG to turbulence 
 has a long history, dating back to the seventies \cite{Forster76,Forster77,Dominicis79,Fournier83},
  and has turned out to be extremely challenging \cite{Adzhemyan99,Zhou10}. 
 In this context, some of us have developed an alternative RG approach, based on non-perturbative and functional RG. 
The NPRG is a modern implementation of Wilson's original idea \cite{wilson74}, which is to calculate the
  physical properties of a system by progressively, scale by scale,  averaging over fluctuations.
 It is an efficient procedure to compute large-scale properties  even in the presence of strong-correlations and 
 fluctuations at all scales (as in critical phenomena) \cite{Berges02}.
 
  The NPRG  consists in constructing a series of  scale-dependent
 effective models, each of which  describing the physics of the system at a given momentum scale $\kappa$.
 An initial condition can be specified when $\kappa$ is a large wave-number scale $\Lambda$, chosen much larger than
 the inverse Kolmogorov scale $\eta^{-1}$, where the dynamics of the velocity field is given by NS equation (or equivalently by
   NS ``bare'' action) \cite{Berges02,Canet16}. 
  The physical statistical properties of the model are obtained
 in the ``infinite volume'' limit $\kappa \to 0$, when all fluctuations have been taken into account.
 
 The \FRG formalism provides  exact RG flow equations,  governing the evolution of these effective models when the
 renormalization scale $\kappa$ runs from $\Lambda$ to 0 \cite{Berges02}. Solving these RG flow equations is thus 
 a way to solve the model.  
 However, these equations are partial-differential and functional equations for the $n$-point
  functions of the theory, that is generalized $n$-point correlation and response functions.
 Moreover, the flow equations for the $n$-point functions involve the $(n+1)$- and $(n+2)$-point functions, 
  such that one should consider in practice an infinite
 hierarchy of flow equations. As common in many theoretical approaches, the usual way to deal with such an
  infinite hierarchy is to devise an approximation scheme to truncate it 
 and obtain closed equation at a given order $n$ (a closure scheme).

  In this work, we focus on two-point ({\it i.e.} two-space point and two-time) functions ($n=2$)
 in the stationary turbulent state.
  More precisely, we consider the scale-dependent 
  correlation function $C_\kappa(t,\vx) = \langle \vv(t,\vx)\cdot \vv(0,0)\rangle_\kappa$ and 
   response function $G_\kappa(t,\vx)$ (translational invariance
   in time and space is assumed), and their Fourier transforms, denoted $C_\kappa(t,\vk)$ and $G_\kappa(t,\vk)$
    where $\vk$ is the wave-vector, and $t$ the time difference in the stationary state.
  The response function 
 is related to the mean value $\langle \vv(t,\vx) \cdot \vf(0,0) \rangle_\kappa$
at scale $\kappa$ through the relation
\begin{equation} 
 \langle \vv(t,\vx) \cdot \vf(0,0) \rangle_\kappa = 2\,(d-1) \,
\int_{\varpi,\vq}  N_\kappa(\vq) e^{-i \varpi  t} G_\kappa(\omega,\vq)\, ,
\label{eq:forcing}
\end{equation}
where $N_\kappa(\vq)$ is the Fourier transform of the correlator of the external stochastic forcing 
 defined in (\ref{eq:correlator}), peaked at scale $\kappa$.  
In particular, one can show that the mean energy injection rate is given by
 \begin{equation}
\label{eq:def-eps}
  \epsilon =  \langle \vv(0,0) \cdot \vf(0,0) \rangle_\kappa = D_\kappa \kappa^{d} \gamma^{-1}\, ,
 \end{equation}
where $D_\kappa$ is the effective (renormalized) forcing strength, $d$ the space dimension, and $\gamma$ 
 a pure (non-universal) number depending on the precise choice of the forcing profile $N(q)$ \footnote{Note that the limit $t\to0$ in
 Eq.~(\ref{eq:forcing}) must be taken with care, as it would naively yield  zero in Ito's prescription, but is in fact non-vanishing, see \cite{Canet16}}.  

\subsection{Closure of the flow equations}

 As mentioned above, the exact NPRG flow equations for the two-point functions involve three- and four-point functions. 
  Usually, in order to solve these equations, one must truncate them in some way.
  Such a truncation (usually called closure) was achieved for the NS problem within the \FRG context in several related works 
   \cite{Tomassini97,Monasterio12,Canet16}. The approximation implemented in these works,
    referred to as Leading Order (LO) approximation, is very much inspired from similar
    ones developed in the context
of the closely related Kardar-Parisi-Zhang (KPZ) equation describing interface growth and kinetic 
roughening \cite{Kardar86}. This approximation scheme has yielded for KPZ very accurate 
results \cite{Canet10,Canet11a,Kloss12},
  and it can be quite straightforwardly transposed to NS since the KPZ and NS 
  field theories share many  common features, in particular (time-gauged)  symmetries 
 \cite{Canet15}.
 
  For NS, the LO approximation consists in  a truncation at quadratic order in the velocity fields 
  (velocity and the associated Martin-Siggia-Rose response velocity), neglecting all higher-order functions
  but the unrenormalized  non-linearity (three-point vertex). This approximation
   is well-controlled in the small wave-number sector $|\vk| \ll \kappa$ of the theory, 
  and thus provides an accurate description of this regime (see detailed discussion in \cite{Canet16}). It was shown at 
 LO that the NPRG flow reaches a fixed point, which encompasses the universal properties of the turbulent  steady state
  \cite{Tomassini97,Monasterio12,Canet16}. 
 The existence of this fixed point was also well-known from perturbative RG \cite{Adzhemyan99,Zhou10}.

  Besides these studies, some of us made a decisive step in \cite{Canet16}, by showing that in the regime of large wave-numbers,
 truncation can be {\it avoided}. 
 The closure of the flow equations for the two-point functions can be achieved without approximation
  in the large wave-number regime, by only
  exploiting the symmetries of the NS action. 
 This result, exceptional  in the \FRG framework,  relies
 on the existence of very constraining symmetries (time-gauged -- or time-dependent, ones,
  in particular a time-gauged shift unveiled in \cite{Canet15}) and the extensive use of the related Ward identities,
 and also on other specificities of NS equation referred to as ``non-decoupling'' (see below).

\subsection{Exact flow equations in the limit of large wave-numbers}

 The equations derived in \cite{Canet16} following from a symmetry-based
  closure are  our starting point. They give  the flow equations for the two-point (response and correlation) functions 
 in Fourier space, and are exact in the limit of large wave-number $|\vk| \gg \kappa$ :
\begin{align}
\kappa \partial_\kappa C_\kappa(\omega, \vk)&=-\frac{2}{3}\, k^2 \, \int_{\varpi}
\frac{C_\kappa(\omega+\varpi, \vk)- C_\kappa(\omega, \vk)}{\varpi^2} \, J_\kappa(\varpi) \nonumber\\
\kappa \partial_\kappa G_\kappa(\omega, \vk)&= -\frac{2}{3} \,k^2 \,\int_{\varpi}
\frac{G_\kappa(\omega+\varpi, \vk)- G_\kappa(\omega, \vk)}{\varpi^2} \, J_\kappa(\varpi) \, ,\label{eq:exactflow-int}
\end{align}
 in units where $\nu=\eta=1$.
 The Fourier conventions used in this work are
\begin{align}
 f(t,\vk) &= \int_{-\infty}^{\infty} \frac{d\omega}{2\pi} \, f(\omega, \vk)\, e^{-i \omega t}\nonumber\\
f(\omega,\vk) &= \int_{-\infty}^{\infty} dt \, f(t, \vk)\, e^{i \omega t} \, ,\label{eq:fourier}
\end{align}
keeping the same notation for the function and its Fourier transform.
 In Eq.~(\ref{eq:exactflow-int}), $J_\kappa$ is the integral
  \begin{align}
J_\kappa(\varpi)
 & =-  \int_{\vq} \Bigg\{2 \p_s N_s(\vq) \,|G_\kappa(\varpi,\vq)|^2 \nonumber\\
 & - 2 \p_s R_s(\vq)\,C_\kappa(\varpi,\vq) \Re \big[G_\kappa(\varpi,\vq)\big]\Bigg\}\, ,
 \label{eq:Ikomega}
\end{align}
where $s=\ln(\kappa/\Lambda)$ is the ``RG time'', with $\p_s = \kappa \p_\kappa$. $N_s$ is the forcing profile defined in (\ref{eq:correlator}), and  $R_s$ is a
 momentum profile involved in the NPRG procedure to freeze all fluctuations with wave-numbers $|\vk| \lesssim \kappa$,
  and which vanishes in the  limit $\kappa \to 0$, see \cite{Canet16} for details.

 As shown in \cite{Canet16},  these flow equations exhibit a very peculiar property  compared to ordinary   
  critical phenomena, named the ``non-decoupling'' property:  the flow does not vanish  
   when the RG  scale is much smaller than   a given wave-number of the correlation function, that is
   \begin{equation}
\lim_{|\vk| \gg \kappa} \frac{\kappa \p_\kappa X_\kappa(\omega,\vk)}{X_\kappa(\omega,\vk)} \neq 0 \, ,
   \end{equation}
where $X_\kappa$ stands for $C_\kappa$ or $G_\kappa$.
    This non-decoupling property opens the door for some intermittency effects.
    Indeed, the stationary turbulent state corresponds to a fixed-point of the flow, which leads 
    to universality and power-law behaviors. However, the non-decoupling property implies that
  the behavior at large wave-numbers can deviate from 
    dimensional scaling, see \cite{Canet16}. This is indeed what we find in the following.
    On the other hand, the decoupling property is restored at {\it equal times}. Indeed, integrating Eq.~(\ref{eq:exactflow-int})
    over frequencies leads to 
    \begin{equation}
\int_{\omega} \kappa \p_\kappa X_\kappa(\omega,\vk) =0\, ,
\label{eq:decoupling}
   \end{equation}  
which means that standard K41 scaling results are recovered at equal times.
 Intermittency corrections are hence absent at leading order in $\vk$ for equal-time correlation and response functions,
  which implies that these corrections must be small, in agreement with experimental and numerical observations \cite{Frisch95}.
 Possible corrections come from the sub-leading orders, which are not included in the  equations studied in this work.
 These corrections would be accessible by integrating the full flow equations, given in \cite{Canet16}, using some approximations to close them, for instance along the lines of Ref.~\cite{Canet11a}.

 The flow equations for the correlation function  and response function in real time can be deduced from Eqs.~(\ref{eq:fourier}) and  (\ref{eq:exactflow-int})
  which yields
\begin{equation}
\label{eq:exactflowt}
 \kappa \partial_\kappa X_\kappa(t, \vk) =-\frac{2}{3}\, k^2 \, X_\kappa(t,\vk) \int_{\varpi} \frac{\cos(\varpi t)-1}{\varpi^2}\,  J_\kappa(\varpi) \, .
\end{equation}
In the following, we consider the regime of small time differences $t \ll \kappa^{-2/3}$, 
 or equivalently large frequencies $\omega \gg \kappa^{2/3}$.
 In this limit,  Eq.~(\ref{eq:exactflowt}) takes a simple form 
\begin{equation}
\label{eq:exactflow}
\kappa \partial_\kappa X_\kappa(t, \vk) =\frac{1}{3}\, k^2 \,t^2 \,I_\kappa\, X_\kappa(t, \vk)\, ,
\end{equation}
where $I_\kappa$ is the integral
\begin{equation}
I_\kappa  =\int_{\varpi} \, J_\kappa(\varpi) \, .
 \label{eq:Ik}
\end{equation}
 Note that
  this integral is a function of the RG scale $\kappa$ only. 
 It does not depend on the external wave-vector $\vk$ and time difference $t$ of the correlation or response functions $X_\kappa$. 
   Moreover, it is determined by the small 
 wave-number sector.
 Indeed, it is shown in \cite{Canet16} that the internal  wave-vector $\vq$ and frequency $\varpi$
 appearing in the integral (\ref{eq:Ik}) are dominated by values of order 
 $|\vq |\lesssim \kappa$ and $\varpi \lesssim \kappa^{2/3}$ respectively,
  which hence belong to the opposite regime as the one studied here. 
  This integral can be reliably (but approximately) computed within an approximation
 controlled in the small wave-number sector, such as the LO one.

\subsection{Fixed point equation}

 As mentioned previously, the RG flow associated to NS turbulence leads to a  fixed-point. 
 We are interested in the vicinity of this fixed point where universality is expected. 
 As usual in RG studies, the fixed point is most conveniently studied in terms of
 dimensionless quantities, such that all explicit $\kappa$-dependence is absorbed. 
 In the \FRG procedure for NS, 
  two renormalized coefficients  $\nu_\kappa$ and $D_\kappa$ are introduced, which correspond respectively
 to the effective viscosity and forcing strength, see \cite{Canet16} for the precise definitions.
  Close to a fixed point, these running coefficients behave
 as power laws $\nu_\kappa \sim \kappa^{-4/3}$ and $D_\kappa \sim \kappa^{-3}$ in $d=3$. 
 The effective viscosity is hence approximately
 related to the microscopic viscosity  as 
\begin{equation}
 \nu_\kappa \simeq \nu (\kappa \eta)^{-4/3} = \epsilon^{1/3} \kappa^{-4/3}\, ,
\label{eq:def-nuk}
\end{equation}
 using Eq.~(\ref{eq:def-eta}) and neglecting the small evolution of $\nu_\kappa$ at the beginning of the flow when it is not a power-law yet. The effective forcing strength $D_\kappa$ is related to the mean injection rate $\epsilon$ following 
Eq.~(\ref{eq:def-eps}). 
 
 We introduce dimensionless variables, denoted with a hat symbol.
  Momenta are measured in units of $\kappa$ and times in units of $(\nu_\kappa \kappa^2)^{-1} 
  = \epsilon^{-1/3} \kappa^{-2/3}$.  The response function $G_\kappa(t,\vk)$  is dimensionless, and the dimension 
  of $C_\kappa(t,\vk)$ is $D_\kappa/(\nu_\kappa \kappa^2) = \gamma \epsilon^{2/3} \kappa^{-11/3}$.
 Let us hence define the dimensionless functions
\begin{align}
 G_\kappa(t,\vk) &=   \hat G_s\left(\hat y = \epsilon^{1/3}\, t k^{2/3}, \hat x = \left({k}/{\kappa}\right)^{2/3}\right)\nonumber \\
C_\kappa(t,\vk) &=  \frac{\gamma \epsilon^{2/3}}{k^{11/3}}  \hat C_s\left(\hat y = \epsilon^{1/3}\,t k^{2/3}, \hat x =\left({k}/{\kappa}\right)^{2/3}\right)\, .
\end{align}
 We also introduce the dimensionless integral $\hat I_s$ as
\begin{equation} 
I_\kappa  = \frac{D_\kappa \kappa^3}{\nu_\kappa \kappa^2} \hat I_s =  \gamma \epsilon^{2/3}\,\kappa^{-2/3}\hat I_s \, .
\end{equation}
According to equation (\ref{eq:exactflow}), the dimensionless functions 
 satisfy the flow equation
\begin{equation}
 \p_s \hat X_s(\hy,\hx)  - \frac{2}{3} \hx \p_{\hx} \hat X_s(\hy,\hx) = \frac{1}{3}\, \hy^2 \, \hx\, \gamma \hat I_s\, \hat X_s(\hy,\hx) \, .
\label{eq:Xfix}
\end{equation}

The fixed point equation corresponds by definition to $\p_s \hat X_s=0$, and  $\hat I_s \to \hat I_*$, $\hat X_s(\hy,\hx) \to \hat X_*(\hy,\hx)$ tend 
to fixed quantities independent of $s$.  Hence, the fixed point equation reads as
\begin{equation}
   \p_{\hx} \hat X_*(\hy,\hx) = -\frac{1}{2} \gamma \hat I_* \hy^2 \,\hat X_*(\hy,\hx)\, \equiv - \hat \alpha \hy^2 \,  \hat X_*(\hy,\hx) \, ,
\label{eq:dif}
\end{equation}
for both functions $\hat C_s$ and $\hat G_s$, with $\hat \alpha$ a non-universal constant (depending on the precise forcing profile.)
  At LO approximation,  one finds that  $\hat I_s$ 
  tends to a positive  constant  $\hat I_*$ of the order 4.10$^{-2}$ at the fixed point,  and  $\hat \alpha$ is hence positive.
   Let us analyze the solution of this equation.

\section{General solution}  
\label{SEC:inertial}

 The general solution of Eq.~(\ref{eq:Xfix}) is given by
\begin{equation}
 \hat X(\hy,\hx) = \, F_X(\hy) \,\exp\left[- \hat \alpha \,\hy^2 \,\hx \right]\, ,
\end{equation}
where $F(\hy)$ is an arbitrary function that must be determined by boundary conditions. 
We are interested in the regime of large wave-numbers $|\vk| \gg \kappa$ and small time differences $t\ll \kappa^{-2/3}$.
 In this regime $\hy \ll t k$. Moreover, the RG flow ensures that all functions are analytic smooth functions.
 Hence, in this regime, $F_X(\hy)$ can be replaced by $F_X(\hy) \simeq F_X(0) \equiv c_X$.

\subsection{Time dependence}

To analyze this solution, let us first restore the dimensions. The physical functions are obtained when the RG scale $\kappa$ tend to 0 (infinite volume limit). In fact, 
  the RG flow reaches a fixed point, and the relevant scale is the inverse integral scale $L^{-1}$: the functions are essentially unchanged when $\kappa$ further decreases. Hence, one has
\begin{align}
 C(t,\vk) &=c_C \, \frac{\gamma\epsilon^{2/3}}{k^{11/3}}\,\exp(-\hat \alpha \epsilon^{2/3} L^{2/3} t^2 k^2)\nonumber\\
   & =  c_C \,\frac{\gamma\epsilon^{2/3}}{k^{11/3}}\,\exp(- \tilde \alpha \epsilon^{2/3} \eta^{2/3} t^2 k^2)\, ,
\label{eq:solC}
\end{align}
and similarly for $G$, 
 with $\tilde \alpha = a \hat \alpha {\rm Re}^{1/2}$ (with $a$  a numerical factor of order one), using that $\eta/L \sim {\rm Re}^{-{3/4}}$ where ${\rm Re}$ is the Reynolds number.  

The expression (\ref{eq:solC}) hence yields that the dominant behavior of the correlation function is a Gaussian dependence in the variable
 $tk$, and not in the scaling variable $tk^{2/3}$.
 This is a non-trivial result, since it implies a violation of standard scale invariance, which is a signature of intermittency.
  It means that the value of the critical
  dynamical exponent $z=2/3$ is effectively  changed to the value $z=1$, 
  which represents a strong correction to Kolmogorov scaling.
 A physical manifestation of this is what is called the sweeping effect, discussed below. 

\begin{figure}
\includegraphics[width=9cm]{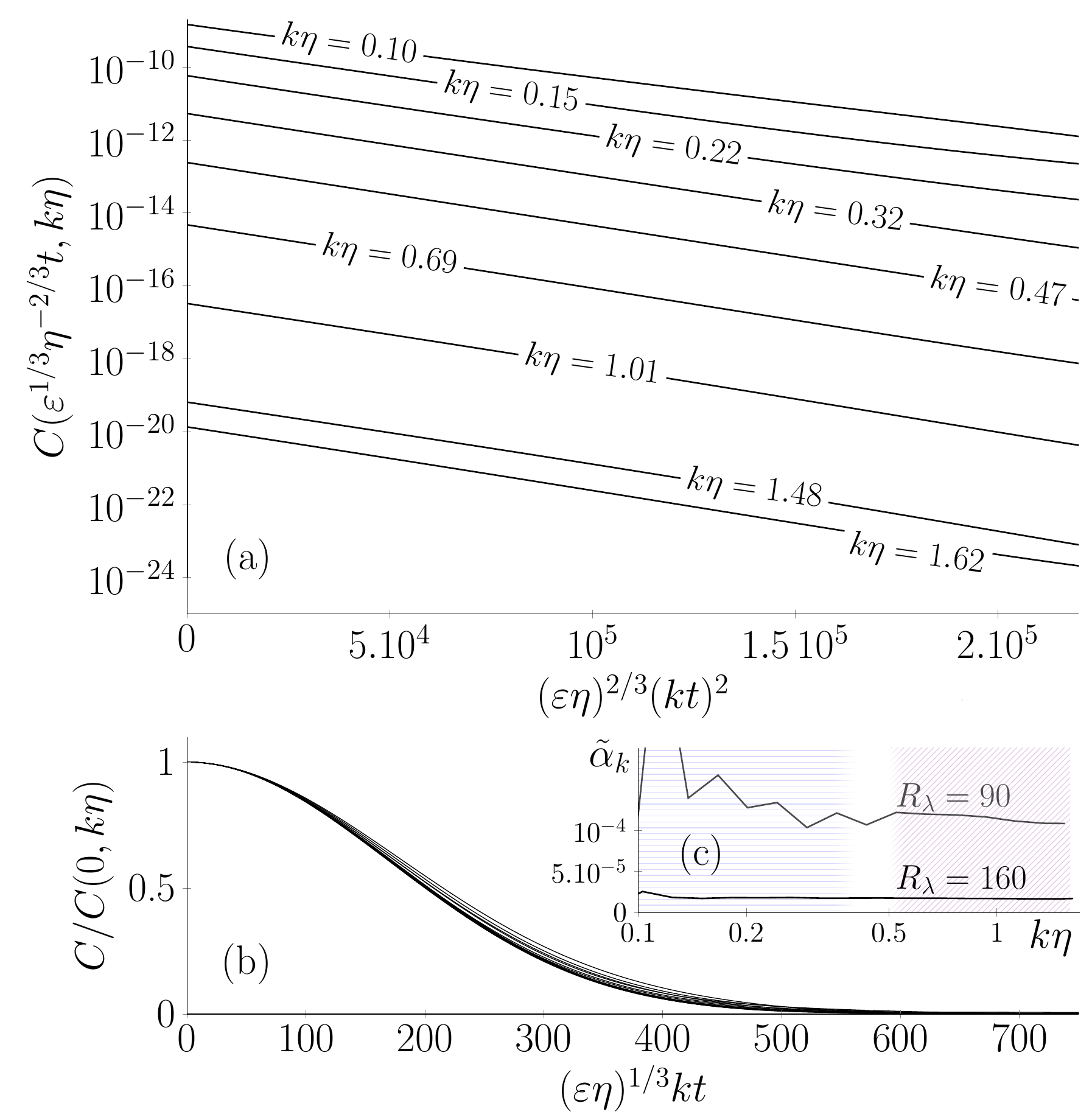}
\caption{(color online) Time-dependence of the correlation function $C(t,k)$ in $k$-space (in dimensionless form) computed
 from direct numerical simulations at $R_\lambda = 90$ and 160.
 (a) Correlation function $C(t,k)$ for $R_\lambda = 160$ in log-scale
 as a function of the dimensionless variable $\hat z^2 = (\epsilon \eta)^{2/3}  (k t)^2$, 
for different values of $k\eta$. In this representation,
 the curves for the different $k\eta$ are parallel lines, with the same slope $\tilde \alpha$,
 confirming the \FRG prediction. (b) Illustration of the collapse
 of the normalised correlation function $C(t,k)/C(0,k)$ for $R_\lambda = 160$ for all values of $k$,
   on a single Gaussian curve in the dimensionless variable $\hat z$.
   (c) Value of the  parameter $\tilde \alpha_k$ estimated from
 gaussian fits of the data for all the different values
 of $k$, for both $R_\lambda = 90$ and 160. $\tilde \alpha_k=\tilde \alpha$ is perfectly constant for $R_\lambda=160$ (approximatively for $R_\lambda=90$).}
\label{fig2}
\end{figure}

 Whereas equal-time quantities are 
easily measured in experiments and numerical simulations, recording the time dependence is more difficult.
 The expression (\ref{eq:solC}) hence provides an interesting prediction, that we tested in numerical simulations, 
  and that could be studied in experimental settings. We performed direct numerical simulations of NS equation at two different Taylor-scale Reynolds number $R_\lambda=90$ and 160.  The detail of the simulation is described in Sec.~\ref{SEC:DNS}.  We computed for both $R_\lambda$ the function $C(t,k)$. The result for $R_\lambda=160$ is represented in Fig.~\ref{fig2}, where $C$
   is plotted as a function of the dimensionless variable $\hat z^2 = \epsilon^{2/3} \eta^{2/3}  k^2 t^2$ 
  for different values of $k\eta$.
  The upper plot is  in log scale. The different curves for each $k\eta$   appear as parallel 
  straight lines as expected from the predicted form (\ref{eq:solC}). 
 The lower plot shows the precise collapse of the normalized function $C(t,k)/C(0,k)$ (in dimensionless form) on a unique Gaussian. 
 The numerical data  hence accurately confirm the \FRG prediction. 
  The values of the parameter $\tilde \alpha$ estimated 
  from the numerical data are given in Table \ref{tab1}. 

\subsection{Inertial range of the energy spectrum}

\begin{figure}
\includegraphics[width=9cm]{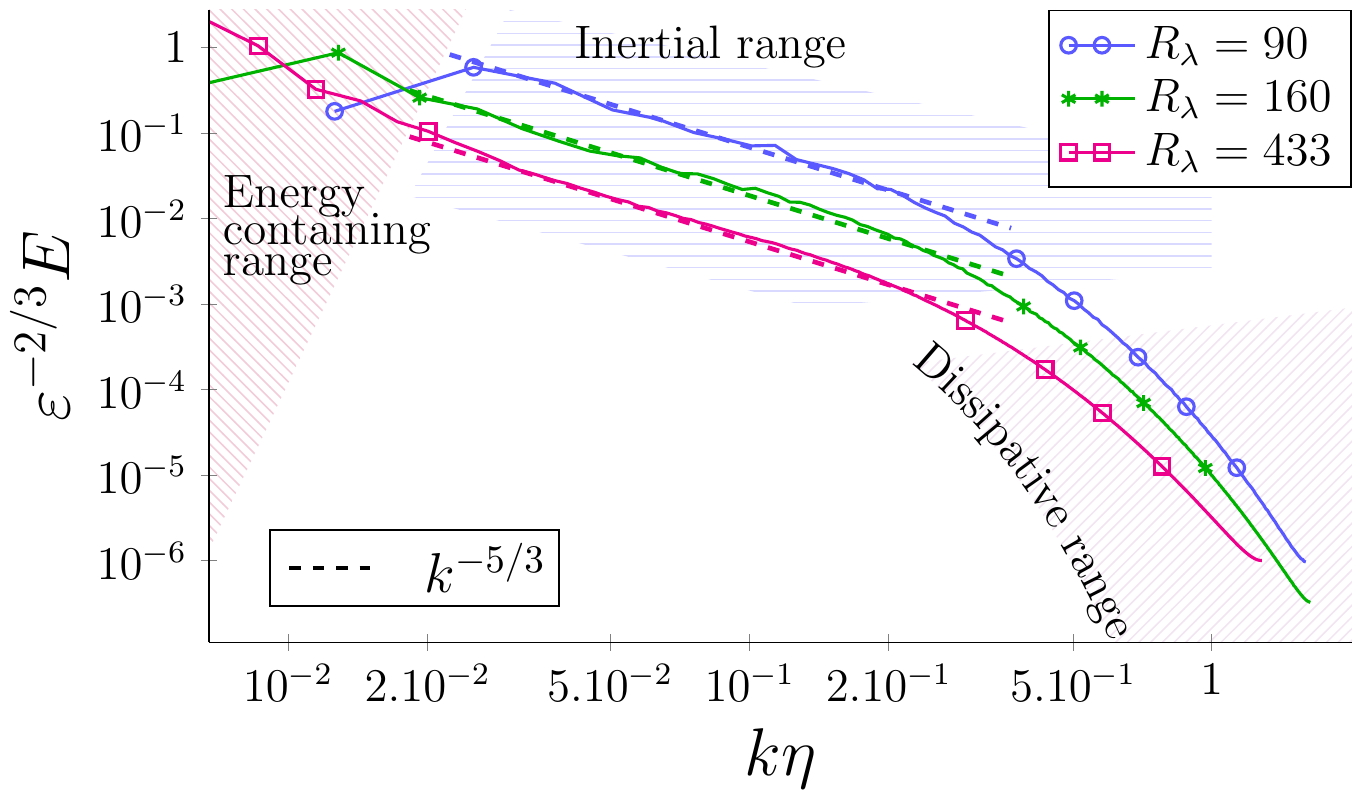}
\caption{(color online) Kinetic energy spectra in $d=3$ obtained from direct numerical simulations at different Taylor-scale Reynolds
 numbers  $R_\lambda=90,160,433$ (from top to bottom), plotted in an dimensionless form : $\epsilon^{-2/3}E$ as a function 
 of $k\eta$. 
 The data for $R_\lambda=90$ and 160 are from this work, the  data for $R_\lambda=433$ come from the John Hopkins
  Turbulence Database \cite{jhtbd}.
  The numerical spectra display an inertial range, extending with $R_\lambda$, with a $k^{-5/3}$ decay
   (up to possible small corrections).}
\label{fig1}
\end{figure}

Equal-time quantities can also be deduced from the solution (\ref{eq:solC}).
 In particular, the kinetic energy spectrum is obtained  as 
\begin{equation}
 E(k) = 4\pi k^2  C(t=0,k) =4\, \pi\, \gamma\, c_C\, \epsilon^{2/3}\,k^{-5/3}\, .
\label{eq:spectrumk}
\end{equation}
It decays as a power-law with the Kolmogorov exponent -5/3. This is in accordance with our statement
  below Eq.~(\ref{eq:decoupling}) that the decoupling property is restored
 at equal times. To illustrate this result, the energy spectra obtained from  numerical simulations
  are shown in Fig. \ref{fig1}, in a dimensionless form. The numerical data come from the simulations we ran
 for $R_\lambda=90$ and $R_\lambda=160$, and from the John Hopkins Turbulence database \cite{jhtbd} for $R_\lambda=433$. 
These spectra exhibit a substantial inertial range,
  extending with $R_\lambda$,
 with a clear $k^{-5/3}$ decay (up to possible small corrections). 

More interestingly, let us  compute
the energy spectrum  as a function of the frequency. This yields
\begin{align}
 E(\omega) &= 4\pi  \int_0^\infty dk\,  k^2 C(\omega,k)  \nonumber \\
           & = 4\pi  \int_0^\infty dk\, k^2 \int_{-\infty}^{\infty} dt \, C(t,\vk)\, e^{i\omega t}\nonumber\\
        &= 2^{8/3} c_C \,\pi^{3/2} \gamma\,\epsilon^{8/9}\,\eta^{2/9}\, \hat \alpha^{1/3}\,  \Gamma(5/6) \,\omega^{-5/3}\, ,
\label{eq:spectrumO}
\end{align}
using Eq.~(\ref{eq:solC}).
Hence,  the decay of the energy spectrum  as a function of the frequency is found to also decay with the exponent -5/3.
 This result corresponds to what is  observed in experiments and numerical simulations  for velocities measured in a fixed 
 reference frame (Eulerian velocities) \cite{Frisch95}.
 The equality of the exponents in (\ref{eq:spectrumk}) and (\ref{eq:spectrumO})  is rooted in the fact that 
 the wave-number and time interval appear as the combination $tk$ in (\ref{eq:solC}), that is with  an effective dynamical exponent $z=1$, 
  and not $z=2/3$, as emphasized previously.  The latter  would  yield
  a $\omega^{-2}$ decay for the energy spectrum in frequency,
 which is characteristic of velocities measured along the flow (Lagrangian velocities), but is not observed for Eulerian
  ones \cite{Chevillard05}.  
Here,  the solution (\ref{eq:solC}) 
  correctly predicts the same power in frequency or wave-number.
 Let us underline that we consider a fluid without mean flow. This observation is hence not related to Taylor's frozen 
 turbulence hypothesis.  In contrast, in the absence of mean flow, the $\omega^{-5/3}$ decay
    is usually attributed to what is named the sweeping effect, which is the random advection of small eddies past the 
    observation point by large energy-containing eddies. This can be viewed as a statistical form of Taylor's hypothesis, 
    as introduced by Tennekes \cite{Tennekes75}. This result is  non-trivial from a field-theoretical point of view, 
    since it implies that standard scale invariance 
 is violated, which can be attributed to intermittency. This shows that this effect is properly taken into account in the solution 
 (\ref{eq:solC}).

\begin{table}[t]
\begin{ruledtabular}
\begin{tabular}{lccc}
  $R_\lambda$           &   90       &   160      &   433    \\ \hline 
 $\lambda$              & 0.236(7)      &  0.161(4)     &   0.118      \\ 
 $\nu$                  &  $10^{-4}$    &   $10^{-4}$   &  $1.8 \times 10^{-4}$\\ 
 $\eta$                 & 0.01264(7)    &  0.00642(4)   & 0.00287 \\
 $\epsilon$             & 0.0000392(9)  &  0.00059(2)  & 0.0928 \\ \hline
 $C_K$                  & 2.26  &  1.90  &  2.24\\
 $\tilde \alpha$               &    1.2$\times 10^{-4}$    &   1.2$\times 10^{-5}$       & --\\
 $\hat \mu$                  &   1.00   &  0.78    & 0.46
\end{tabular}
\end{ruledtabular}
\caption{\label{tab1} Parameters of the numerical simulations for the different  Taylor Reynolds number $R_\lambda$: 
Taylor micro-scale $\lambda$, viscosity $\nu$,  Kolmogorov scale $\eta$, mean energy injection rate $\epsilon$;
 and fitting parameters: Kolmogorov constant $C_K$,  dimensionless parameter of the Gaussian time-dependence $\tilde \alpha$ 
 and dimensionless parameter of the
  stretched exponential decay $\hat \mu$,  see text.}
\end{table}

\subsection{Dissipative range of the energy spectrum}
\label{SEC:dissipative}

The solution (\ref{eq:solC}) is valid for large wave-numbers and small time intervals. Let us study more precisely the limit of 
 vanishing time-differences.  In the previous section, $t$ was sent strictly to 0. This supposes in turn 
 that one can consider arbitrary small time differences, which
   is equivalent to sending Kolmogorov scale $\eta$ to 0. This is of course justified in the inertial range, but not on smaller scales,
 that is in the dissipative range. In practice, the smaller time difference is bounded by Kolmogorov time
\begin{equation}
 \tau = \left(\frac{\nu}{\epsilon}\right)^{1/2}\, .
\end{equation}
  Let us hence consider
 the limit  $t\to \tau$, and $k \gg \kappa \sim L^{-1}$. It is reasonable to assume that the scaling variable $\hy$
  saturates in this limit to a constant value given by
\begin{equation}
\hy = \epsilon^{1/3} t k^{2/3} \to   \hy_0 = \epsilon^{1/3} \tau L^{-2/3} = \eta^{2/3} L^{-2/3} \, .
\end{equation}
 The fixed point equation (\ref{eq:dif}) then becomes
\begin{equation}
  \p_{\hx} \hat X_*(\hy \to   \hy_0,\hx) =  - \hat \alpha \eta^{4/3}\, L^{-4/3} \,  \hat X_*(\hy,\hx)\, .
\end{equation}
The general solution of this equation reads
\begin{equation}
 \hat X_*(\hy\to \hy_0,\hx) = c_X \,\exp(-\hat \alpha  \eta^{4/3} L^{-4/3} \,\hx)\,
\end{equation}
and the energy spectrum is thus given in this limit by a stretched exponential 
 \begin{equation}
E(k)  = 4\pi \gamma \epsilon^{2/3}\, k^{-5/3}  c_C  \exp\left[- \hat \alpha  \eta^{4/3} L^{-2/3} k^{2/3}  \right]  \, .
\end{equation}
Remarkably, a new scale emerges in the exponential, which is  Taylor scale. 
Indeed, the latter is related to the Kolmogorov and integral scale through 
\begin{equation}
 \frac{\lambda}{\eta} \sim {\rm Re}^{1/4}\, ,\quad \hbox{and} \quad\frac{\lambda}{L} \sim {\rm Re}^{-1/2} \, ,
\end{equation}
from which one deduces that
\begin{equation}
 \frac{\eta^{2/3}}{L^{1/3}} \sim \lambda^{1/3}\,{\rm Re}^{-3/4} ,
\end{equation}
and thus 
 \begin{equation}
E(k)  = 4\pi \gamma \epsilon^{2/3}\, k^{-5/3}  c_C  \exp\left[- \hat \mu  (\lambda k)^{2/3}  \right]  \, ,
\label{eq:exp}
\end{equation}
with $\hat \mu$ a non-universal constant  $\hat \mu = b \hat \alpha {\rm Re}^{-3/2}$, where $b$  a numerical factor of order one.
\begin{figure}[t]
\includegraphics[width=9cm]{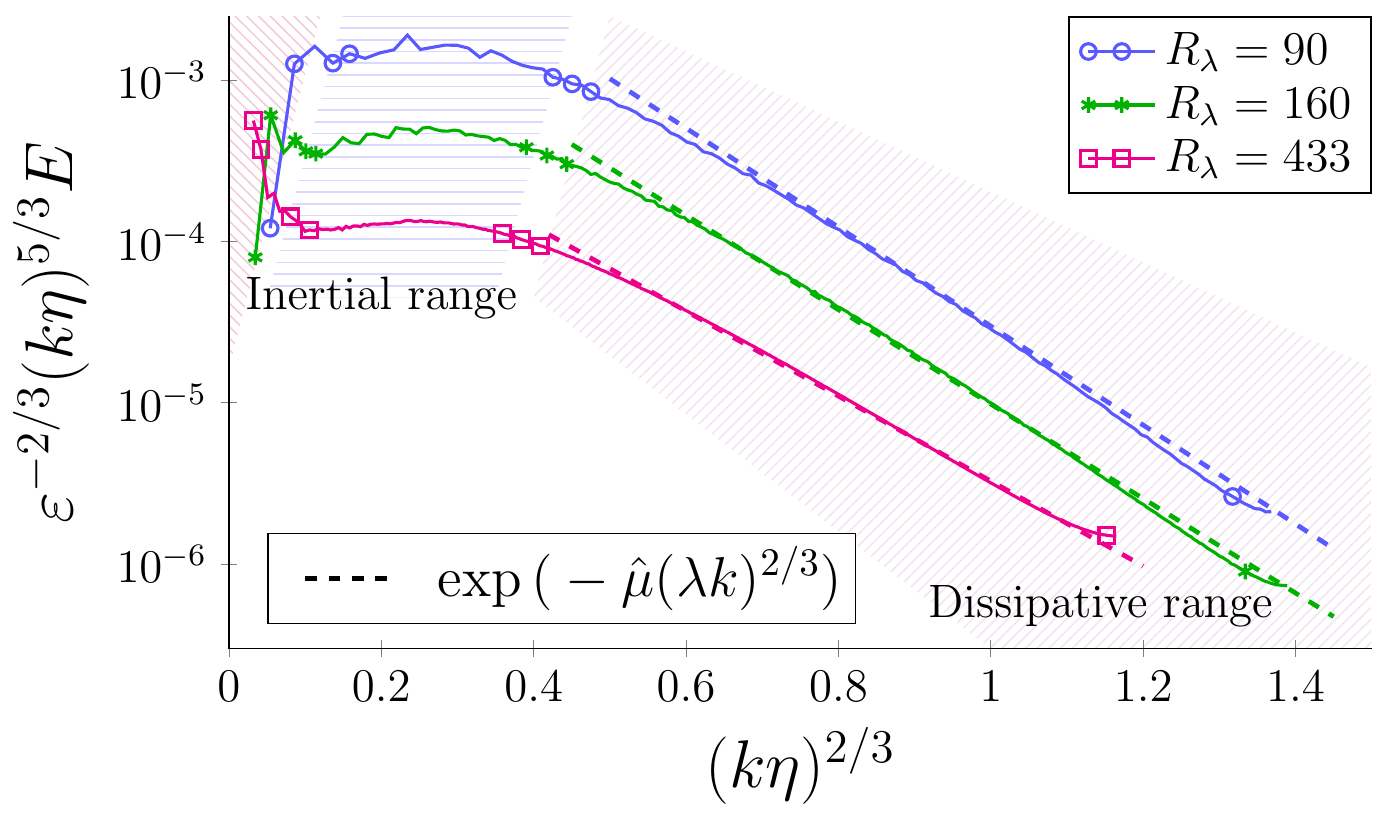}
\caption{(color online) Same dimensionless kinetic energy spectra as in Fig. \ref{fig1}, multiplied by $(k\eta)^{5/3}$
 and represented in log-scale as a function of $(k\eta)^{2/3}$. 
The  \FRG predicts a crossover from the $k^{-5/3}$ power-law to  a stretched exponential decay $\exp(-\hat \mu (\lambda k)^{2/3})$ in the dissipative range (dashed lines), which is  observed in the numerical data (plain lines with symbols).}
\label{fig3}
\end{figure}
Hence, the expression (\ref{eq:exp}) shows that, beyond the $k^{-5/3}$ Kolmogorov decay, a crossover to a stretched exponential decay with
 argument $k^{2/3}$ occurs typically below the Taylor scale, that is for wave-numbers  in 
  the dissipative range. Several expressions have been proposed for this regime, mainly under the form of a modified 
  exponential
  $\exp(-c k^{y})$, but with different values for $y$ (1/2 \cite{Tatarskii67}, 3/2 \cite{Uberoi69}, 4/3 \cite{Pao65} or 
  2 \cite{Novikov61,Gurvich67}). These expressions are mostly based on approximate fits of the experimental data 
 or (approximate) analytical considerations \cite{Monin73}. The common wisdom is that the spectrum decay is ``approximately 
 exponential'' in the dissipative range \cite{Nelkin89}.
 In order to assess the prediction (\ref{eq:exp}), 
  we analyzed the numerical energy spectra in the dissipative range. The  stretched 
  exponential on the scale $k^{2/3}$ is indeed observed, as illustrated in Fig. \ref{fig3},
  although the number of decades available in the data is of course limited. 
 The value of the parameter $\hat \mu$ estimated from the numerical data is given  in Table~\ref{tab1}.

\section{Direct Numerical Simulations}
\label{SEC:DNS}

\begin{figure}[h]
\includegraphics[width=6cm]{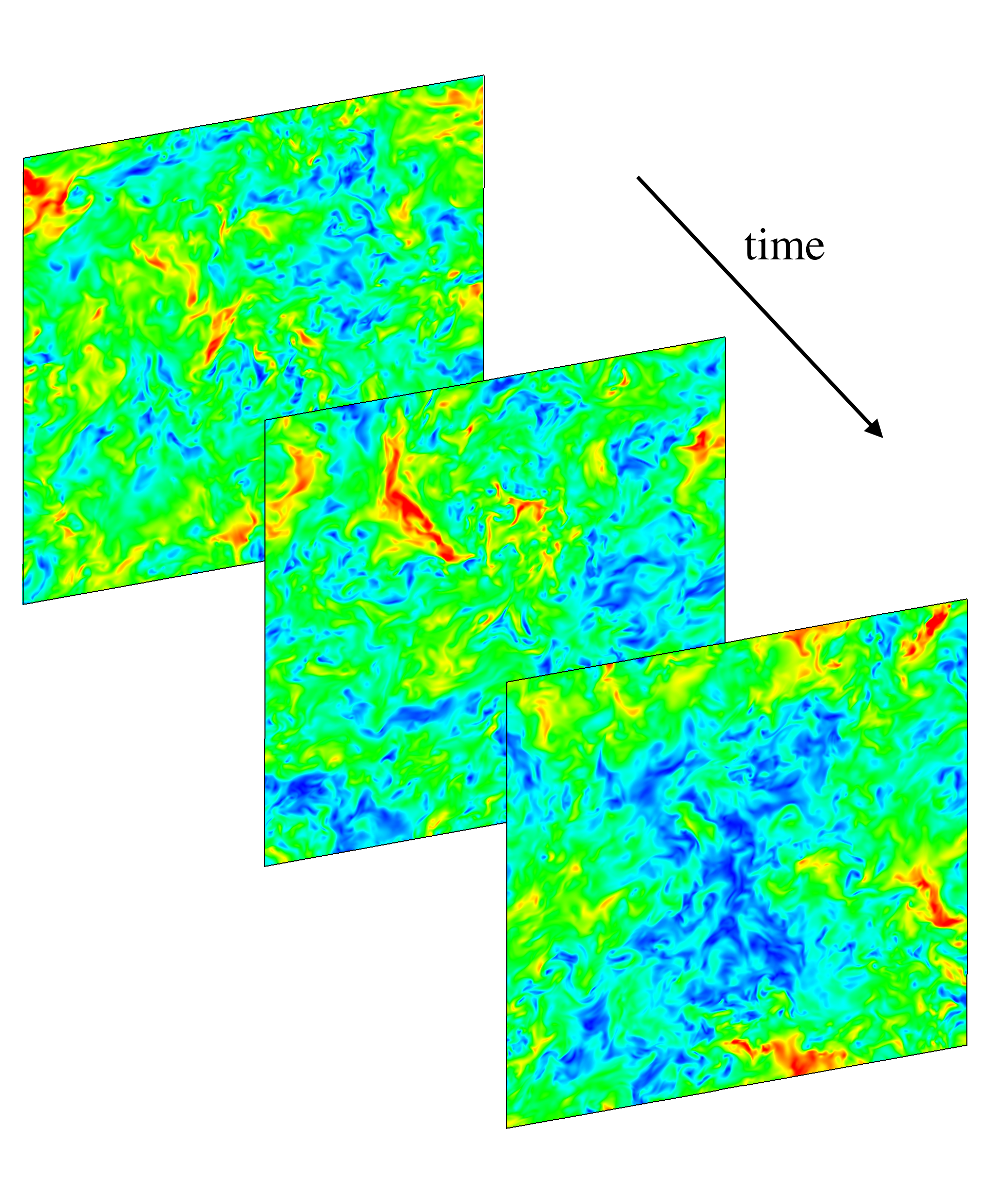}
\caption{(color online) Typical map of the modulus of the velocity field in simulated turbulence for $R_\lambda=160$, at different times.}
\label{fig4}
\end{figure}
The numerical simulations performed to obtain the data  of incompressible forced homogeneous isotropic turbulence
 at $R_\lambda=90$ and 160 are based on a pseudo-spectral code  with second-order explicit Runge-Kutta time-advancement
  \cite{Lagaert14}. The simulation domain is discretized using $256^3$, resp. $512^3$, grid points on a domain of 
  length $2\pi$ for $R_\lambda=90$, resp. $160$. A classic 3/2 rule is used for dealiasing the non-linear convection 
  term and a projection method in spectral space is used to enforce the divergence-free condition. The forcing is a 
  fully random forcing concentrated at small wave numbers \cite{Alvelius99}. The simulation parameters are chosen such
   that $k_{\rm max} \eta> 1.5$, where $k_{\rm max}$ is the maximum wavenumber in the domain. 
 A typical configuration of the modulus of the velocity field obtained in the simulation for $R_\lambda=160$ is represented 
 at different times in Fig. \ref{fig4}.
 Additional data from the John Hopkins Turbulence Database \cite{jhtbd} are also used
 for the kinetic energy spectra in Fig.~\ref{fig1} and \ref{fig3}. They correspond 
 to simulations  on 1024$^3$ nodes of isotropic turbulence with $R_\lambda \simeq 433$. The relevant parameters 
 for the simulations are gathered in Table \ref{tab1}.

\section{Conclusion}

In this work, we provide an analytical expression for the space and time dependent correlation (and response) functions
 of a fluid in a  forced turbulent state. These expressions are derived from \FRG flow equations
 which are exact in the limit of wave-numbers large compared to the inverse integral scale.
 We show that these expressions yield predictions beyond the standard observations and Kolmogorov theory.
   The essential aspects are i) the time-dependence for the correlation function 
 in $k$-space  $\propto\exp(-\alpha t^2 k^2)$ with an effective dynamical exponent $z=1$, which implies a strong
 correction to standard scaling theory; ii)  a related $\omega^{-5/3}$ decay 
 of the energy spectrum, as observed for Eulerian velocities, and which reflects the sweeping effect; 
  iii)  a stretched exponential decay 
 as $k^{-5/3}\exp(-\hat \mu (\lambda k)^{2/3})$ of the spectrum in the dissipative range. We believe that deriving such 
 analytical solutions, directly from NS equation, and not on phenomenological basis, constitutes a major progress
 in the theoretical understanding  of isotropic and homogeneous turbulence, and its modelling at all scales. 

It opens new perspectives in many respects.
 An important issue to be addressed is the numerical integration of the complete
 flow equations (in both the small and large wave-number sectors),  to assess intermittency corrections for equal-time quantities.
 This analysis would require to make some approximations in order to truncate the full flow equations.
  Another important direction is the investigation of two-dimensional turbulence, and the derivation of correlation
   functions for scales both below the integral scale (direct cascade) and above (inverse cascade).
 Moreover, a promising perspective  is the computation of higher-order structure functions, and the determination of 
 intermittency effects in this case, which are expected  to be much more pronounced as the order increases.

\begin{acknowledgements}
The authors thank B. Delamotte, L. Chevillard and M. Tarpin for fruitful discussions.  Simulations were performed 
 using HPC resources from GENCI-IDRIS (Grant 020611).
\end{acknowledgements}

\bibliographystyle{prsty}
%\bibliography{../../Biblio-KPZ/biblioKPZ}

\end{document}